\documentclass[11pt,a4paper,twocolumn,english]{article}
\usepackage[T1]{fontenc}
\usepackage[utf8]{inputenc}
\usepackage{amsmath}
\usepackage{graphicx}
\usepackage{setspace}
\usepackage{amssymb}
\usepackage[numbers]{natbib}

\makeatletter

\usepackage{float}
\usepackage{changepage}

\usepackage[multidot]{grffile}

\makeatother

\usepackage{babel}

\begin{document}
\newcommand{\seq}{B}
\newcommand{\conf}{\mathbf{G}}
\newcommand{\dd}{\mathrm{d}}
\newcommand{\gfor}{\mu}
\newcommand{\kT}{k_{\text{B}}T}
\newcommand{\FE}{A}
\newcommand{\fe}{a}

\title{DNA nano-mechanics: how proteins deform the double helix}

\author{Nils~B.~Becker\thanks{Corresponding author. Address: Labo de Physique de l'ENS, 46 all\'ee de l'Italie, 69007 Lyon, France}
\\Laboratoire de Physique de l'\'Ecole Normale Sup\'erieure,\\
Universit\'e de Lyon, France 
\and Ralf~Everaers\\Laboratoire de Physique de l'\'Ecole Normale Sup\'erieure,\\
Universit\'e de Lyon, France }

\maketitle
\begin{abstract}
\begin{singlespace}
It is a standard exercise in mechanical engineering to infer the external
forces and torques on a body from a given static shape and known elastic
properties. Here we apply this kind of analysis to distorted double-helical
DNA in complexes with proteins: We extract the local mean forces and
torques acting on each base-pair of bound DNA from high-resolution
complex structures. Our analysis relies on known elastic potentials
and a careful choice of coordinates for the well-established rigid
base-pair model of DNA. The results are robust with respect to parameter
and conformation uncertainty. They reveal the complex nano-mechanical
patterns of interaction between proteins and DNA. Being non-trivially
and non-locally related to observed DNA conformations, base-pair forces
and torques provide a new view on DNA-protein binding that complements
structural analysis.\end{singlespace}

\end{abstract}
\newpage{}

\section*{Introduction}

A large class of DNA-binding proteins induce deformations of the DNA
double helix which are essential in biochemical processes such as
transcription regulation, DNA packing and replication \citep{garcia07}.
Insight into the mechanism of binding largely depends on high-resolution
structures of DNA-protein complexes. A first step in their analysis
consists of a description of DNA conformation in the complex,  often
in terms of a suitably reduced set of degrees of freedom such as the
rigid base-pair parameters, e.g.~\citep{richmond03}. As a second
step, sites of local DNA deformation can be identified by comparison
with ensembles of fluctuating DNA conformations. This allows to quantify
deformation strength in terms of a free energy. Here we take the analysis
a step further by extracting the points of attack, magnitudes and
directions of forces acting between protein and DNA in the complex. 

The basic idea of inferring the force on an elastic body from its
deformation is as commonplace as stepping on a scale to measure one's
weight. We propose to apply the same idea to DNA-protein complexes,
using DNA as a nanoscale force probe calibrated by a known elastic
potential. That is, starting from the coarse-grained mean conformation
of a piece of bound DNA as extracted from a high-resolution structural
model, we infer the corresponding coarse-grained static mean forces
required for that conformation. To implement this idea, we use the
rigid base-pair level of coarse-graining. Correspondingly, our analysis
results in a DNA base-pair step elastic energy profile, complemented
by the set of mean forces and torques by which the protein acts on
each DNA base-pair.

This article focuses on the theoretical basis, implementation, range
of applicability and validation of DNA nano-mechanics analysis. We
begin by discussing the statistical mechanics of the mechanical equilibrium
in DNA-protein complexes in section Background. We also motivate our
choice of the rigid base-pair level of coarse-graining which, unlike
standard molecular mechanics with atomistic force fields, allows reliable
extraction of mean forces within the experimentally available resolution.
Our matrix formalism for force and torque calculations is described
in section DNA nano-mechanics, and implementation and parameter choice
details are given in Methods. In the Results section, we present exemplary
force and torque calculations for several high-resolution NMR and
x-ray complex structures. These examples show the robustness of the
analysis with respect to experimental and parameter uncertainties,
and demonstrate the key features of base-pair forces and torques described
in the Discussion section: Base-pair forces and DNA deformation are
nontrivially and non-locally related, and they allow to discriminate
force-transmitting and non-transmitting protein-DNA contacts. Based
on these features, DNA nano-mechanics analysis has a number of promising
applications, such as validation and design of coarse-grained molecular
models for multi-scale simulations, and identification of target sites
for structure-changing mutations in protein-DNA complexes. These are
expanded upon in the Conclusion section.

\section*{Background\label{sec:Background}}

\begin{figure}
\begin{centering}
\includegraphics[width=0.42\textwidth]{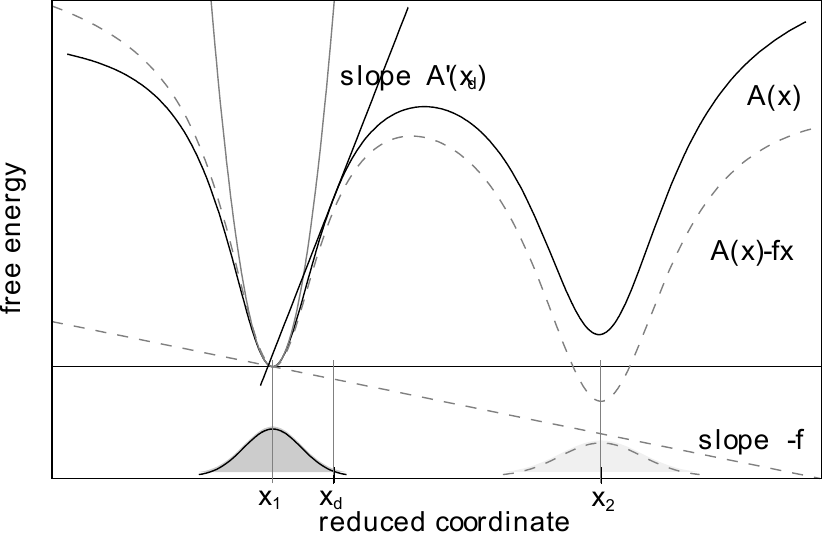}
\par\end{centering}

\caption{Constraint force $f_{d}$ and externally applied force $f$ in a stereotyped
double-well free energy landscape, and thermal distribution (solid
lines). Under an external force $f$, the landscape is tilted (dashed
lines).}
\label{fig:doublewell}
\end{figure}

The statistical mechanics of DNA can be described on multiple levels
of coarse-graining, depending on the required amount of detail. For
a chosen set of reduced coordinates $\{x\}$, the corresponding free
energy $A(x)$ is a potential of mean force, i.e.~constraining the
system to a conformation $x_{d}$ requires the mean force $f_{d}=A'(x_{d})$
. This holds regardless whether or not $A$ is approximately quadratic;
for instance, $A$ could have a shape as in Fig.~\ref{fig:doublewell}. 

From high-resolution structures of protein-DNA complexes one obtains
the mean conformation $x_{d}$ within some uncertainty $\delta_{d}$,
and the size $B^{1/2}=\langle(x-x_{d})^{2}\rangle^{1/2}$ of thermal
fluctuations around it. If the force is approximately linear over
the range $B^{1/2}$ , i.e.~if \begin{equation}
A'''<2A'/B,\label{eq:fluctineq}\end{equation}
then the mean force by which the environment acts upon DNA to produce
the observed conformation is given as $f_{d}=A'(x_{d})\pm A''(x_{d})\delta_{d}$.
This simple scheme fails for atomistic force fields since they are
strongly nonlinear on the $B^{1/2}\gtrsim5\,\text{\AA}$, violating
Ineq.~\ref{eq:fluctineq}. Mean atomistic forces would have to be
extracted from a full MD simulation with adequately constrained average
(not instantaneous) positions. In contrast, mean forces acting on
groups of atoms may be meaningfully extracted from coarse-grained
descriptions of a single structure, when the smoother coarse-grained
free energy is compatible with Ineq.~\ref{eq:fluctineq}. 

Here we consider the rigid base-pair \citep{calladine82} model of
DNA as a good compromise between resolution and reliability. The corresponding
sequence-dependent free energies have been parametrized from microscopic
data \citep{lankas03,olson98} with considerable effort. They have
been successfully used in describing indirect readout \citep{ahmad06,becker06,morozov05,steffen02}
and match well with known, $\mu$m scale DNA elastic properties \citep{becker07}.
The free energy functions, and therefore the extracted forces, are
reliable within some sufficiently sampled region around the ideal
B-DNA ground state which excludes only the most extreme deformations,
see Discussion. Notably, the sampled free energy within this region
is well approximated by a quadratic function \citep{olson98,lankas03},
leading to linear elastic forces and validity of Ineq.~\ref{eq:fluctineq}.
While extensions of the free energy function into the anharmonic region
and inclusion of trinucleotide coupling could further improve on  the
range of validity and accuracy base-pair forces, the present parametrization
leads to consistent results, as shown below. Note that by choosing
the rigid base-pair level of coarse-graining, all information on force
pairs that cancel on smaller scales, e.g.~separation of bases, is
disregarded. The resulting description retains those forces that are
relevant for large scale DNA deformations.

\section*{DNA nano-mechanics\label{sec:DNA-nano-mechanics}}

\begin{figure}
\begin{centering}
\includegraphics[width=0.32\textwidth]{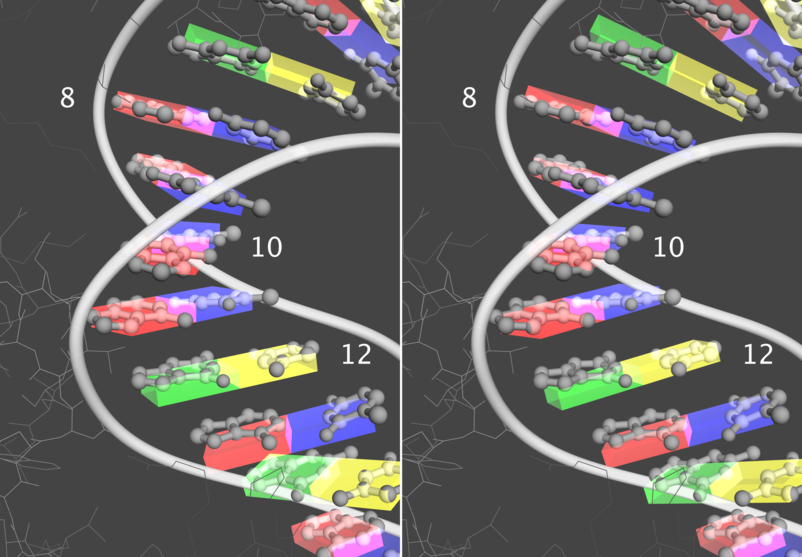}
\par\end{centering}

\caption{Rigid base-pair model. Atomic coordinates are reduced to the base-pair
center position and orientation degrees of freedom, represented as
bricks. Conformations of the I\emph{ppo}-I complex before (left) and
after (right) pre-relaxation within a range of $r_{m}=0.3\mathring{\mathrm{A}}$,
see Methods.}
\label{fig:relaxbox}
\end{figure}

In the rigid base-pair model of DNA \citep{calladine82}, base-pairs
are represented by as rigid bodies without internal structure, see
Fig.~\ref{fig:relaxbox}. From the atomic coordinates of a base-pair
$k$, a reference frame $\mathbf{g}_{k}$ is derived in a standardized
way \citep{olson01,lu03}, which specifies the base-pair orientation
$\mathbf{R}_{k}$ and position in space $\mathbf{p}_{k}$, both given
relative to some fixed lab frame. The conformation of a piece of DNA
is described by the chain $\conf=(\mathbf{g}_{1},\mathbf{g}_{2},\dots,\mathbf{g}_{n})$
of base-pair reference frames. The base-pair \emph{step }conformations
are denoted by $\mathbf{g}_{k\, k+1}$, i.e.~the orientation and
position of $\mathbf{g}_{k+1}$ relative to $\mathbf{g}_{k}$. The
data $\mathbf{R}_{k}$ and $\mathbf{p}_{k}$ may be represented in
a number of different coordinate systems, including Euler angles,
exponential coordinates, etc. Similarly, the relative step conformations
$\mathbf{g}_{k\, k+1}$ are conventionally discussed in terms of the
six base-pair step parameters Tilt, Roll, Twist, Shift, Slide and
Rise \citep{dickerson89}. At the moment, we avoid fixing a particular
coordinate system, considering the frames $\mathbf{g}_{k}$ as abstract
elements of the rigid motion group.

Unusual DNA conformations can be recognized by their low probabilities
in an equilibrium ensemble of freely fluctuating DNA. The corresponding
conformational elastic free energy $\FE_{B}(\mathbf{G})$ depends
on the base sequence of the chain $B=b_{1}b_{2}\dots b_{n}$. The
free energy $\FE_{B}$ is a potential of mean force which quantifies
the strength of deformation of the chain. We fix the zero-point so
that $\min_{\mathbf{G}}\FE_{B}=0$ and call $\FE_{B}$ the elastic
energy of the chain as is customary. 

We argue that knowledge of the function $\FE_{B}$ can provide valuable
information when it is used to derive mean forces%
\footnote{By `forces' we always mean mean forces in the following.%
}: The elastic generalized force by which the chain acts on its $k$-th
base pair is given by the negative derivative of the chain energy
with respect to the $k$-th base pair configuration, $-\dd_{\mathbf{g}_{k}}\FE_{\seq}(\conf)$.
In the static equilibrium of a DNA-protein complex, this elastic force
exerted by DNA is balanced by the \emph{external force} acting on
base-pair $k$\begin{equation}
\gfor_{(k)}=\dd_{\mathbf{g}_{k}}\FE_{\seq}(\conf).\label{eq:genbal}\end{equation}
The basic force balance Eq.~\ref{eq:genbal} allows to infer external
forces from DNA conformations. This relation holds for general elastic
energy functions, including next-nearest neighbor coupling \citep{yanagi91,packer00a,beveridge04,fujii07}.
However, the best presently available full parameter sets approximate
the elastic energy as a sum of harmonic, nearest-neighbor base-pair
step energies $\fe_{bb'}$ so that \begin{equation}
\FE_{B}(\mathbf{G})=\sum_{k=1}^{n-1}\fe_{b_{k}b_{k+1}}(\mathbf{g}_{kk+1}).\label{eq:pairpot}\end{equation}
For details on our particular choice, see Methods. In this case Eq.~\ref{eq:genbal}
specializes to\begin{equation}
\mu_{(k)}=\dd_{\mathbf{g}_{k}}\fe_{b_{k-1}b_{k}}(\mathbf{g}_{k-1k})+\dd_{\mathbf{g}_{k}}\fe_{b_{k}b_{k+1}}(\mathbf{g}_{kk+1}).\label{eq:forcebal}\end{equation}
One sees that the external force on a base-pair $\mathbf{g}_{k}$
balances a sum of two terms, which are just the elastic \emph{tensions}
in the steps $k-1,k$ and $k,k+1$, respectively.%
\footnote{Eq.~\ref{eq:forcebal} is the equivalent of the standard relation
`force = div stress' of continuum elasticity in the present context
of a discrete, linear chain.%
} At each end of the chain, there is of course only one-sided tension.

Unlike the generalized force itself, the components $\mu_{(k)i}$
are defined with respect to a particular choice of basis. We pick
a basis by requiring that the $\mu_{(k)i}$ have simple physical interpretations
in terms of force and torque.%
\footnote{This requirement precludes the use of Euler angles for the orientation
$\mathbf{R}_{k}$.%
} To formulate this idea, we remark that the frames $\mathbf{g}_{k}$
are elements of the rigid motion group and can be represented as so-called
homogeneous matrices, which are well-known in robotics, see~\citep{murray}.
This approach has been used for coarse-graining the rigid base-pair
model~\citep{becker07} and in the context of worm-like chain models
of DNA~\citep{chirikjian00}. Here, base-pair frames are written
explicitly as $\mathbf{g}_{k}=\left[\begin{smallmatrix} & \mathbf{R}_{k} &  & \mathbf{p}_{k}\\
0 & 0 & 0 & 1\end{smallmatrix}\right]$ where $\mathbf{R}_{k}$ is a $3\times3$ rotation matrix and $\mathbf{p}_{k}$
is a $3\times1$ column vector. The base-pair step conformations are
calculated as a matrix product $\mathbf{g}_{kk+1}=\mathbf{g}_{k}^{-1}\mathbf{g}_{k+1}$
from the base-pair conformations, where $\mathbf{g}_{k}^{-1}=\left[\begin{smallmatrix} & \mathbf{R}_{k}^{\mathsf{T}} &  & -\mathbf{R}_{k}^{\mathsf{T}}\mathbf{p}_{k}\\
0 & 0 & 0 & 1\end{smallmatrix}\right]$. The corresponding matrix generators $\mathbf{X}_{i}$ for rotations
($1\leq i\leq3$) and translations ($4\leq i\leq6$), are $4\times4$
matrices with entries $(\mathbf{X}_{i})_{jk}=\epsilon_{jik}+\delta_{k4}\delta_{i-3j}$.
Using this notation, the derivatives of $\FE_{B}$ with respect to
infinitesimal motions of base-pair $k$,\begin{equation}
\mu_{(k)i}=\tfrac{d}{dh}\bigl|_{0}\FE_{B}(\mathbf{g}_{1},\dots,\mathbf{g}_{k-1},\mathbf{g}_{k}(\mathbf{1}+h\mathbf{X}_{i}),\mathbf{g}_{k+1},\dots,\mathbf{g}_{n}),\label{eq:lidiff}\end{equation}
have the required simple interpretations: $(\mu_{(k)i})_{1\leq i\leq3}$
are the Cartesian components of the external \emph{torque} $\mathbf{t}_{(k)}$
on base-pair $k$ about an axis through $\mathbf{p}_{k}$, while $(\mu_{(k)i})_{4\leq i\leq6}$
are the Cartesian components of the external \emph{force} $\mathbf{f}_{(k)}$
attacking at $\mathbf{p}_{k}$. These components are relative to
the base-pair fixed triad $\mathbf{R}_{k}$. For actually calculating
the components Eq.~\ref{eq:lidiff}, it is convenient to rewrite
the step energy $\fe_{bb'}$ which is usually given in terms of the
rigid base-pair step parameters, in terms of exponential coordinates,
see Supplementary Material, \ref{sub:expcoords}.%
\begin{figure}
\begin{centering}
\includegraphics[width=0.42\textwidth]{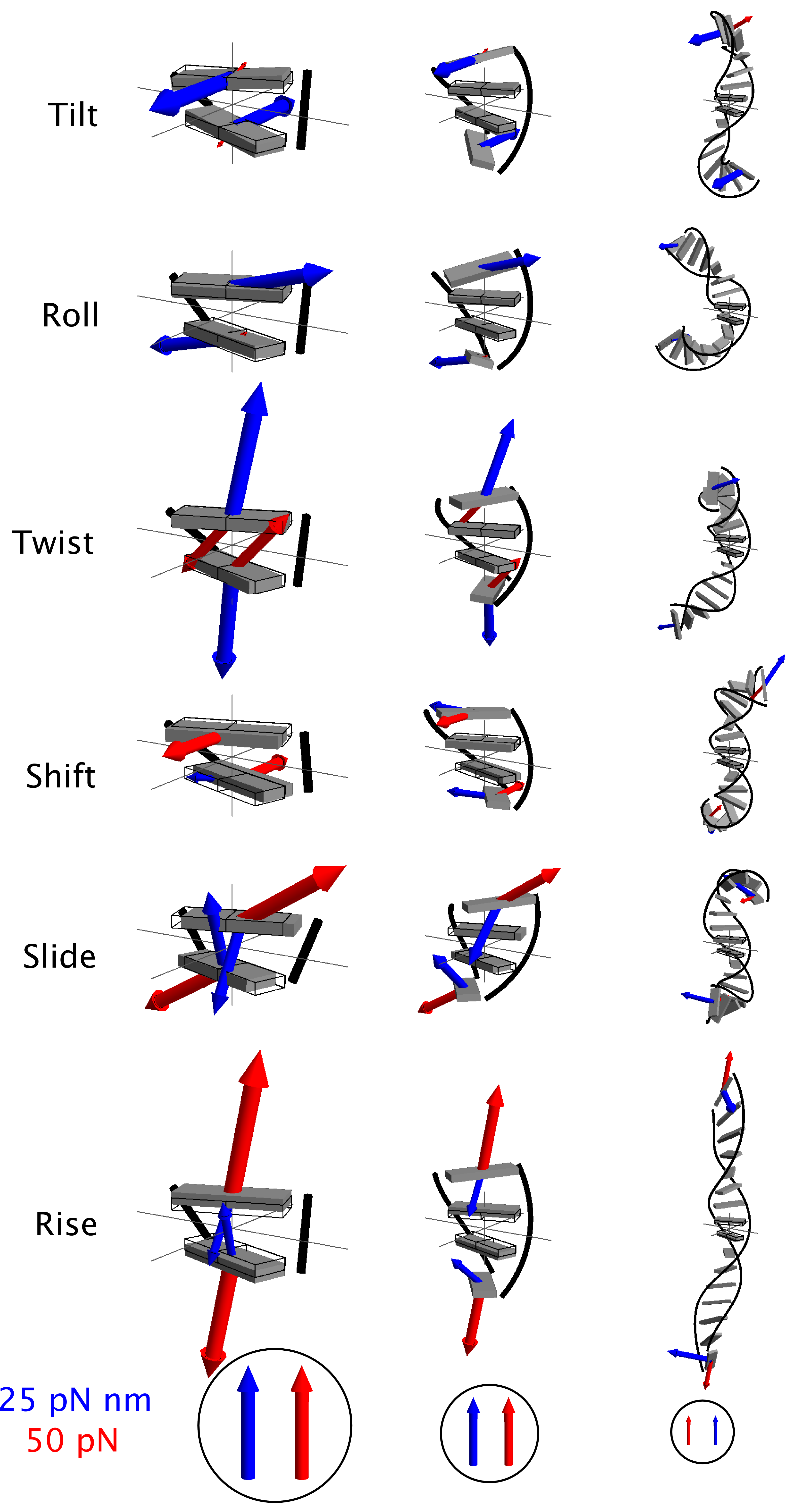}
\par\end{centering}

\caption{Force and torque pairs acting on DNA to produce an excess of one base-pair
step parameter in each row. Torque vectors $\mathbf{t}_{(1)},\mathbf{t}_{(2)}$
shown in blue, force vectors $\mathbf{f}_{(1)},\mathbf{f}_{(2)}$,
in red. The same deformation of the \emph{central} base-pair step
can be produced by external force and torque pairs attacking directly
(left column), at the nearest neighbor base-pairs (middle column),
or seven base-pairs away (right column). Sequence-averaged MP parameters.
For plots of the base-pair step parameters associated with these equilibrium
shapes, see Fig.~\ref{fig:portplots}.}
\label{fig:portraits}
\end{figure}

An overview of the relation between external forces and torques, and
base-pair step deformations is given in Fig.~\ref{fig:portraits}.
Consider first the left-hand column. The force and torque pairs required
for increasing each of the six base-pair step parameters demonstrate
the strong coupling between the different base-pair deformation modes
of B-form DNA. E.g.~to produce pure increased Twist, overwinding
torque must be assisted by compressive forces as a consequence of
the counter-intuitive twist-stretch coupling \citep{lionnet07,lionnet06,gore06}.
In addition, there exists a geometric coupling effect: Force balance
requires that force pairs sum to zero. However, they need not be collinear;
any offset of their lines of attack generates an additional torque
`by leverage' which enters the torque balance. Thus the external torque
vector pairs shown in Fig.~\ref{fig:portraits} do generally not
sum to zero.

A base-pair step can be deformed by external forces acting directly
on its constituent base-pairs, but also indirectly, by external forces
at distant base-pairs. Examples of this non-local effect are shown
in the middle and right hand columns of Fig.~\ref{fig:portraits}.
Here the central base-pair step of each chain has the same deformation
as in the left hand column; however this time produced indirectly,
by external forces applied only at the chain ends. Along the chain,
tensions are non-zero but balanced so that DNA assumes a stressed
equilibrium shape \citep{coleman03} in which all intermittent $\mu_{(k)}$
vanish. These shapes can exhibit strongly non-uniform deformation,
e.g.~non-uniform Twist, see Fig.~\ref{fig:portplots}. (For a related
study of stress localization in RNA, see~\citep{fernandez94}.) Given
these complicated shapes, it is difficult to guess at external forces
by structural inspection.

In the general case of DNA bound to protein, external forces may act
anywhere along the chain. Here each base-pair step deformation is
caused by a combination of local external forces and internal propagated
tension. In this article, instead of investigating equilibrium shapes
for given boundary conditions, we focus on the converse question of
what local external forces and torques are required for a given general
shape. These forces and torques give a quantitative measure for how
the proteins forces DNA into that shape.

\section*{Methods\label{sec:Methods}}

\subsection*{Rigid base-pair parameter sets}

To parametrize the base-pair step energy $\fe_{bb'}$, step equilibrium
conformations $\mathbf{g}_{\text{eq},bb'}$ and stiffness matrices
$\mathbf{S}_{bb'}$ are needed for each dinucleotide type $bb'$.
We used a combination of knowledge-based and simulation-based parameters.
Specifically, the relaxed conformation of each dinucleotide $bb'$
is the mean conformation reported for $bb'$ in a crystal structure
database of protein-bound DNA fragments \citep{olson98}. The stiffness
matrix $\mathbf{S}_{bb'}$ for each dinucleotide is the stiffness
matrix reported for $bb'$ as a result of explicit-solvent, all-atom
MD simulations of a library of oligonucleotides \citep{lankas03}.
We have reported previously on the the relation of this hybrid parameter
set to pure knowledge-based or pure simulation based parameters \citep{becker06},
and have shown that it reproduces known values $\mu$m-scale elasticity
of DNA without fitting \citep{becker07}. In these articles and here,
the hybrid parameter set is denoted by `MP'. To evaluate the robustness
of our method, we have compared the results computed with MP to those
computed with a pure knowledge-based parameter set `P'. The P parameter
set is essentially identical to the P$\cdot$DNA parameter set from
\citep{olson98} constructed from mean values and covariance matrices
of protein-bound DNA \citep{olson98}, the difference being a global
multiplicative factor to correct the temperature scale, see \citep{becker06}.
The dependence of our results on the choice of parameter set is discussed
in Results.

\subsection*{Restrained Relaxation}

To account for the high, but limited precision of structural models,
we included an initial pre-relaxation stage, a strategy which has
been used also for atomistic force fields in related studies \citep{paillard04b,morozov05}.
Here, the rigid base pair coordinates of the DNA fragment were allowed
to relax simultaneously, descending the gradient of $\FE$; at the
same time, each base-pair was restrained to a region around its original
conformation by a sharply increasing potential. We set the size $r_{m}$
of this region on the order of the atomic position uncertainty. In
this way elastic tension in DNA is allowed to relax, but only so little
that the relaxed conformation remains consistent with the reported
structural data. The parameter $r_{m}$ describes the assumed precision
of the input data, and is not a property of the employed force field.
To set $r_{m}$ we used the estimated coordinate error based on a
Luzzati plot as reported in the PDB files if available (which is mostly
around 15\% of the reported x-ray resolution), and 15\% of the resolution
otherwise. This gave a range of $r_{m}=0.28\dots0.33\mathring{\mathrm{A}}$
in the shown examples. The effect of this relaxation on base-pair
frame conformations is barely visible by eye, see Fig.~\ref{fig:relaxbox}.

The pre-relaxation procedure can be seen as a form of data smoothing
with a tendency to equalize elastic tension in DNA. Force analysis
after restrained relaxation produces the set of \emph{smallest} external
forces which are compatible with the confidence region of the structural
input data. In practice, the effect of relaxation was to reduce extreme
peaks in the resulting energy and force profiles, while relaxing the
weakest external forces to zero. Thus, relaxation reduces extreme
force outliers and eliminates low-level random noise. 

Although any sensible choice of $r_{m}$ should roughly equal the
structural uncertainty, its exact value remains undetermined. The
global scale of our computed forces and torques depends on the value
of $r_{m}$, but their relative magnitudes along the chain and their
directions are only weakly affected by this choice. After relaxation,
we also found substantially increased agreement between energies and
forces computed using different elastic parameter sets. These features
of pre-relaxation are illustrated by Fig.~\ref{fig:relax}.

\subsection*{Implementation and Visualization}

Forces and torques in this article were computed starting from the
following Protein Data Bank \citep{bernstein77} high-resolution structures:
\textsf{1nvp}, \textsf{1cdw}, \textsf{1qne} for TBP, \textsf{1l1m}
for lac repressor and \textsf{1cz0} for I\emph{ppo}-I. Rigid base-pair
frames were computed from atomic coordinates by least-squares fitting
of model base-pairs following a standardized procedure, using the
3DNA program \citep{olson01,lu03}. Calculation of energies, forces
and torques as described in section DNA nano-mechanics, as well as
pre-relaxation were implemented in Mathematica. Three-dimensional
vector depictions of base-pair conformations, forces and torques were
exported as VRML files, which are available as Supplementary Material
Data S1-S3. They can be visualized and superimposed with atomic structure
data using molecular visualization software, for instance the free
molecular visualization program Chimera \citep{pettersen04} which
was used for the images in this article.

\section*{Results\label{sec:Results}}

We have applied an analysis as described above for three x-ray co-crystal
structures of TATA-box binding protein \citep{nikolov96,patikoglou99,bleichenbacher03}
and an ensemble of 20 NMR solution structure conformers of a lac repressor
complex \citep{kalodimos02}. As an example of a trapped intermediate
state of a structural modification of  DNA, we analyzed a co-crystal
structure of the homing endonuclease I-\emph{ppo}I \citep{galburt99}.%
\begin{figure*}[t]
\begin{centering}
\includegraphics[width=0.95\textwidth]{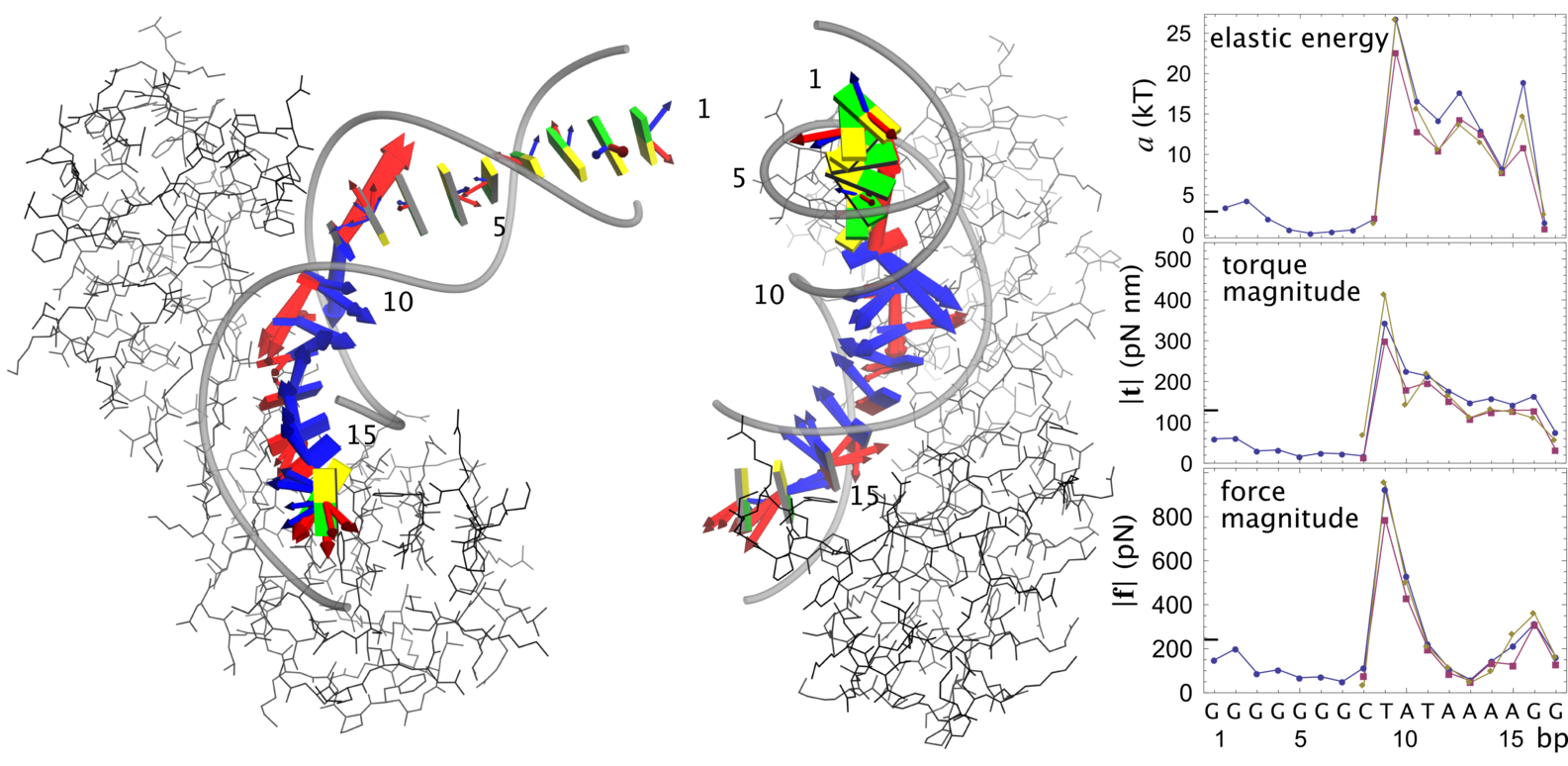}
\par\end{centering}

\caption{TFIIA(not shown)-TBP-DNA complex with force and torque vectors from
three different TBP-DNA crystal structures, left. Corresponding energy,
force magnitude and torque magnitude profiles, right. Here and in
the following figures, linear forces $\mathbf{f}_{(k)}$ are shown
as red arrows, torques $\mathbf{t}_{(k)}$ as blue arrows; base-pairs
are represented as numbered small boxes with sequence coloring, `A'
red, `T' blue, `G' green,`C' yellow; the two viewpoints are rotated
by $90^{\circ}$ around the vertical axis. Sequence (5', base-pair
1)-GGGGGGGCTATAAAAGG-(3', base-pair 17). Allowed relaxation range
$r_{m}=0.3\mathring{\mathrm{A}}$ in all complexes. MP parameter set.
The three-dimensional representations of base-pairs, force and torque
vectors used for this figure are available, as detailed in Methods
(Supplementary Material, Data S1).}
\label{fig:tbp}
\end{figure*}

Fig.~\ref{fig:tbp} illustrates the force analysis of TATA-box binding
protein (TBP) complexed with cognate DNA. In this complex, TBP bends
DNA into the major groove. The overall turn of about $80^{\circ}$
is distributed over the eight base-pairs of the TATA-box, whose steps
have uniform positive roll. The highest deformation energy occurs
at the first `TA' dinucleotide step whose base-pairs (9 and 10) are
separated but not strongly kinked. A secondary peak in elastic energy
can be seen at base-pair step 15-16. Inspection of the structure shows
that at both of these locations, a phenylalanin residue partially
intercalates between the base-pairs. The initial, straight poly-G
DNA region shows only small deformation energies. 

Superimposed on the TFIIA-TBP-DNA crystal structure \citep{bleichenbacher03},
Fig.~\ref{fig:tbp} shows force and torque vectors from an analysis
of three different co-crystals of TBP with the same DNA binding site.
A strong opposing force pair is seen to pull apart base-pairs 9 and
10. Along the box, the $80^{\circ}$ turn is associated with a nicely
aligned sequence of torque vectors at base-pairs 10 to 15; they deviate
by at most $28^{\circ}$ from pointing into the major groove. Unlike
the rather evenly distributed torques, the base-pair forces have a
minimum in the center of the box, and a second peak associated with
a force pair stretching the base-pair step 15-16. Note that the directions
of forces and torques at base-pairs 9-10 and 15-16 are approximately
related by a two-fold symmetry around step 12-13 corresponding to
the symmetry of TBP; however their magnitudes are about half at base-pairs
12-13. Base-pairs 1-8 are present in only one of the crystal structures,
and exhibit low force and torque magnitudes%
\begin{figure*}[t]
\begin{centering}
\includegraphics[width=0.95\textwidth]{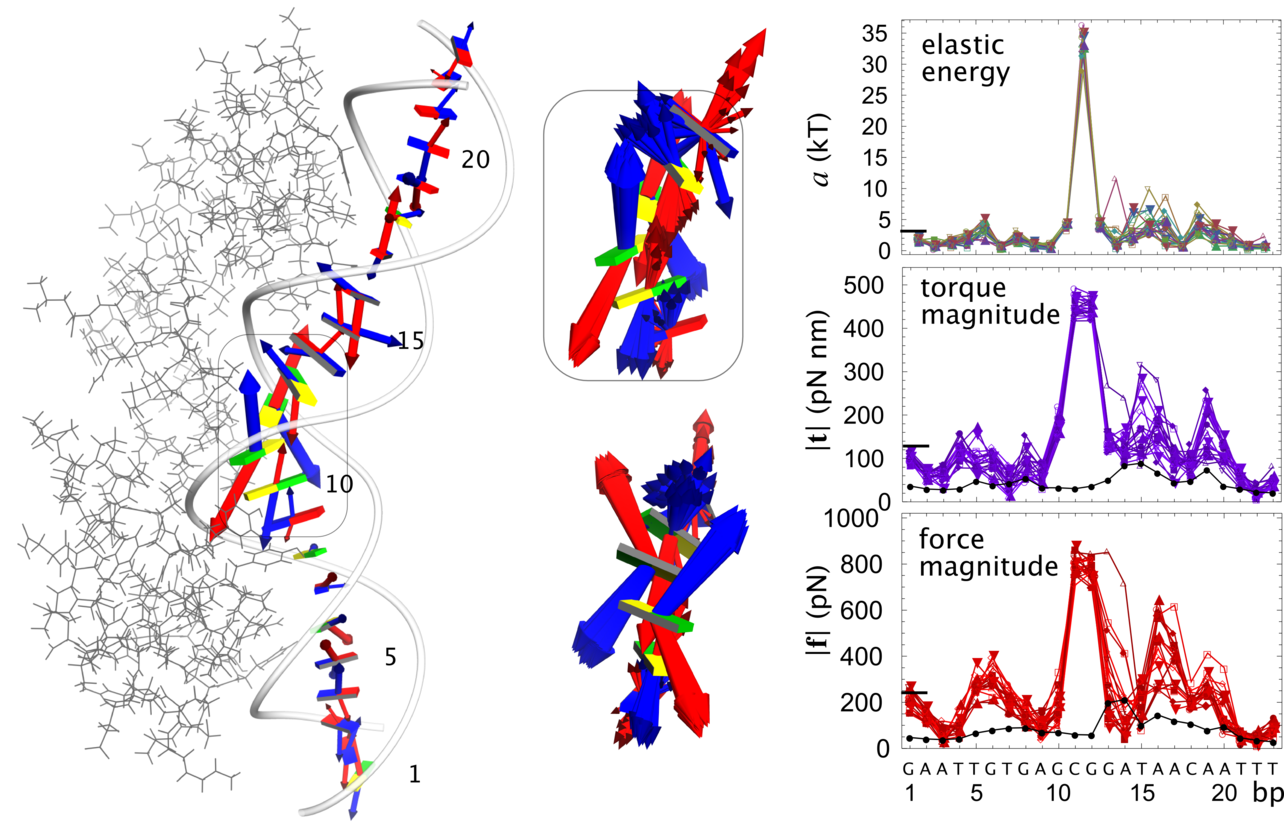}
\par\end{centering}

\caption{Complex of lac repressor and DNA. NMR structure of one out of 20 conformers
with ensemble mean force and torque vectors, left. Two magnified views
of the encircled region, middle column, with force and torque vectors
of all conformers. Ensemble energy, torque and force magnitude profiles,
right; the ensemble standard deviation profiles $\langle|\mathbf{f}_{(k)}|^{2}\rangle_{\text{NMR}}^{1/2}$
and $\langle|\mathbf{t}_{(k)}|^{2}\rangle_{\text{NMR}}^{1/2}$ of
force and torque vectors are shown in black. MP parameter set. Sequence
(5', base-pair 1)-GAATTGTGAGCGGATAACAATTT-(3', base-pair 23). The
three-dimensional representations in Data S2.}
\label{fig:nmr}
\end{figure*}

Fig.~\ref{fig:nmr} shows the nano-mechanics analysis of an NMR solution
structure ensemble of \emph{E. coli }lac repressor bound to DNA. In
this complex, a wild-type operator with non-palindromic sequence is
bound by a homodimer of the lac repressor DNA binding domain \citep{kalodimos02}.
The resulting complex structure is only approximately two-fold symmetric.
The strongest deformations occur around the central six base-pairs
9-14, producing the overall $30^{\circ}$ bend of DNA. The kink at
the symmetry center 11-12 has by far the highest elastic energy. 

As in the TBP case, the peak of elastic energy is associated with
a pair of base-pair stretching forces, here accompanied by a strong
unwinding torque pair. In the central region 9-14, force and torque
directions closely follow the approximate two-fold symmetry of the
complex, despite the fact that the sequence is asymmetric. Outside
the central region, force and torque symmetry is broken. Secondary
peaks in torque and force magnitude can be identified at the symmetry-related
base-pair steps 6-7 and 16-17. At base-pairs 6 and 7, a rather weak
pair of shearing forces attack, while base-pairs 16 and 17 are pulled
apart by a strong stretching force pair. Also, a strong torque on
base-pair 19 has no counterpart at the symmetry-related position 4,
highlighting the different binding modes of the two half-sites.%
\begin{figure*}[t]
\begin{centering}
\includegraphics[width=0.95\textwidth]{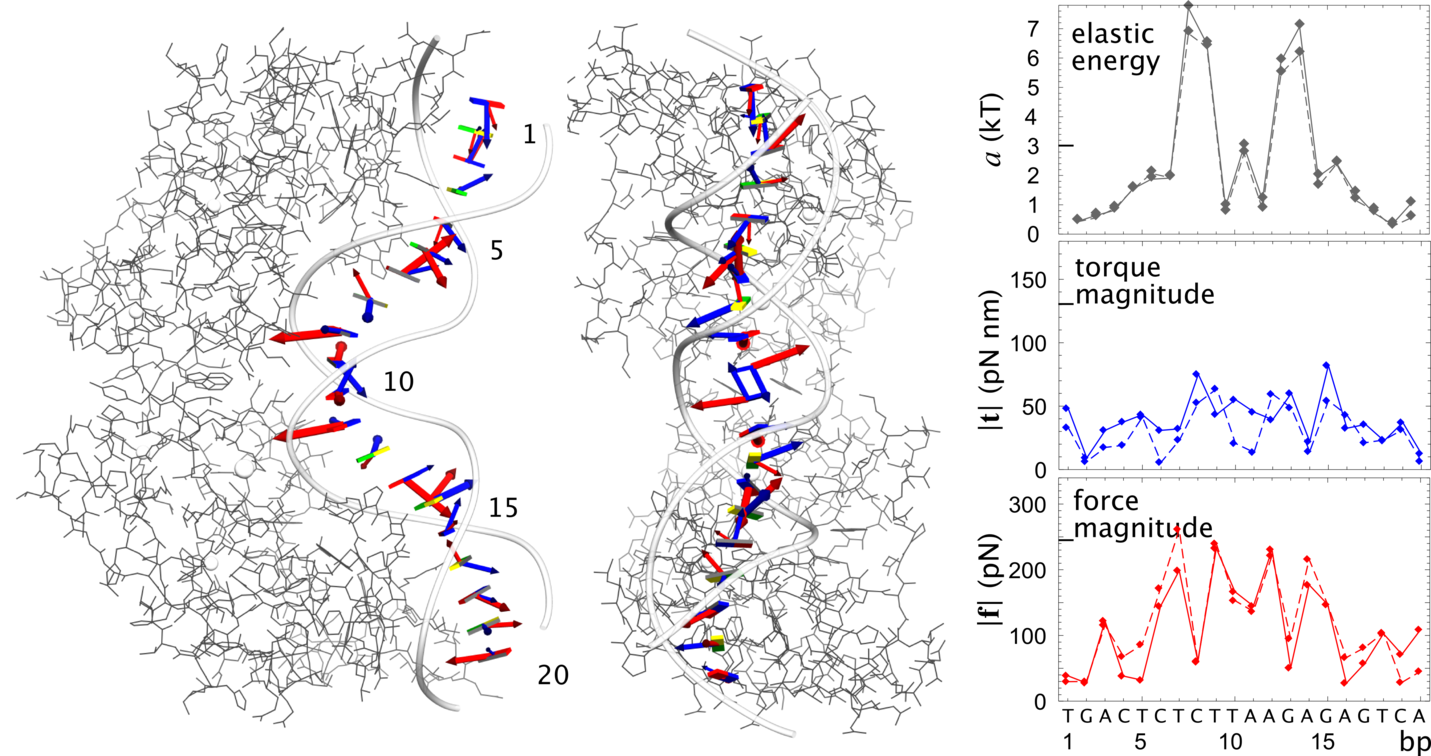}
\par\end{centering}

\caption{I\emph{ppo}-I DNA complex. The points of single-strand cuts in the
functional complex are indicated. Relaxation range $r_{m}=0.3\mathring{\mathrm{A}}$.
The MP parameter set was used for vectors, and MP (solid line) and
P (dashed line) parameter sets for profiles, right. Sequence (5',
base-pair 1)-TGACTCTCTT$\cdot$AAGAGAGTCA-(3', base-pair 20). Three-dimensional
representations in Data S3.}
\label{fig:ippo}
\end{figure*}

Fig.~\ref{fig:ippo} illustrates the analysis of the homing endonuclease
I\emph{ppo}-I, bound to target DNA substrate in an un-cut state. This
complex has a palindromic operator sequence and an overall two-fold
symmetry. Cleavage occurs within at step 8-9 (and the symmetry-related
12-13) in the active form of the complex. In contrast to lac, deformation
of the operator occurs mainly not at the symmetry center 10-11 but
within the triplet 7-8-9 (12-13-14). The intervening base-pair steps
are sheared and tilted, producing the overall $45^{\circ}$ bend of
the binding site. 

The computed external forces show that these deformation are mainly
due to a pair of strong opposing forces attacking at base-pairs 7
and 9 (12 and 14), stretching and shearing the triplets. In addition,
there is an external torque attacking at the intermediate base-pair
8 (13). While the adjacent base-pair step 9-10 (11-12) is almost completely
relaxed, the central step is sheared by an opposing lateral force
pair.

\section*{Discussion\label{sec:Discussion}}

We point out the main general features of an analysis of DNA-protein
complex structures in terms of base-pair forces and torques, using
the structures presented in the Results section as examples.

\subsection*{Robustness of nano-mechanics analysis}

The conformational data on which our analysis is based, as well as
the elastic parameter sets, are reliable within certain bounds of
error. How do these sources of uncertainty influence the derived external
forces and torques?

To assess the dependence on details of crystallization, we computed
forces and torques based on three x-ray structures of the TBP complex,
see Fig.~\ref{fig:tbp}. Two of the complexes lack TFIIA and all
three crystals have different space groups and thus different crystal
contacts. Nonetheless, their energy, force and torque profiles quantitatively
agree at common base-pairs. As can be seen from the three-dimensional
representation, also force and torque directions agree closely.

The conformational variability within an NMR structure ensemble leads
to variability of elastic energies, external forces and torques. These
were computed for a lowest-energy ensemble of 20 conformers of the
lac repressor solution structure \citep{kalodimos02}, see Fig.~\ref{fig:nmr}.
We find that the main features of the corresponding profiles are clearly
more pronounced than the variation across the ensemble. Also forces
and torque directions are robust among conformers, with the exception
of the most weakly forced base-pairs.

When comparing computed forces and torques corresponding to the different
parameter sets P and MP, we find surprisingly good agreement (Fig.~\ref{fig:ippo}),
considering the completely different sources (crystal structure database
for P, MD simulation for MP, see Methods) of the stiffness parameters. 

The choice of pre-relaxation range $r_{m}$ reflects the assumed precision
of the structural input. It does not strongly affect the relative
magnitudes of local forces and torques. However a present limitation
of the method comes from the fact that overall force and torque scales
vary with $r_{m}$. Setting $r_{m}$ to the structural precision,
leads to the lowest external forces that are compatible with the considered
structural model. While this choice is reasonable, it could be improved
upon by incorporating information on the equilibrium fluctuations
of bound base-pairs derived from the local B-factors. Clearly, a direct
comparison to averaged forces from full simulations of protein-DNA
complexes would be enlightening. Note however that compared to the
construction of hybrid base-pair potentials \citep{becker06}, the
correction of known artifacts such as systematic undertwist is less
straightforward for atomic force fields.

\subsection*{Force scale and validity range}

The characteristic scales of base-pair forces and torques in thermal
equilibrium at room temperature are determined by equipartition of
energy. They result as 245~pN and 130~pN~nm, respectively. (For
base-pair step tensions one obtains 170~pN and 90~pN~nm.) Thus
in a thermal equilibrium ensemble, the instantaneous forces of a harmonic
base-pair step are normally distributed with width $245\,\text{pN}$,
so that on average 10~\% of instantaneous forces are higher than
$400\,\text{pN}$. We conclude that our elastic potentials are well
supported by MD simulation up to around $400\,\text{pN}$ and $200\,\text{pN\, nm}$.

In our examples, only a few of the highest force peaks exceed that
range. This can be expected also for most other complexes: In the
DNA-protein crystal structure database \citep{olson98}, the bulk
of DNA base-pair steps was slightly \emph{less} deformed \citep{becker06}
than in the thermal ensemble \citep{lankas03}. Note however that
outlying base-pair steps do occur in the crystal database; they are
preferentially deformed into the softest directions \citep{olson98}.
Together with the observation of unexpectedly frequent sharp bending
of DNA \citep{wiggins06} this suggests that the true free energy
function stays below the harmonic approximation for strong deformations,
cf.~Fig.\ref{fig:doublewell}. Thus forces and torques tend to be
overestimated outside the validity range given above. 

Our force field is parametrized by MD simulations at room temperature,
while the crystal structures are typically observed at $\simeq110\,\text{K}$.
The computed forces are thus predictions for the complex at room temperature
under the assumption that the mean conformation is essentially the
same as at $110\,\text{K}$. This assumption is ubiquitous in structural
biology when interpreting crystal structures in terms of biological
function. On the other hand, we cannot assume that the mean conformation
from the crystal, or the force field, are valid at temperatures that
approach the melting temperature of DNA. Not only will the increased
thermal fluctuations exceed harmonic range of the potential described
above. Also, the stiffness parameters themselves change with temperature
due to the entropic part contained in the elastic free energy. Finally,
the basic requirement of the rigid base-pair model that internal degrees
of freedom are uncoupled between neighboring base-pairs is violated
by cooperative base-pair opening. As a result, it is hard to estimate
the temperature range of validity of the present force calculation.
A very loose upper bound is $\simeq330\,\text{K}$ where local bubble
formation starts.

The limitations listed above arise from the presently available free
energy functions; the general procedure of force and torque extraction
is unchanged for general anharmonic free energies, or free energies
that include inter-base degrees of freedom.

\subsection*{Protein forces may exceed critical forces for DNA structural transitions}

Even though within thermal range, forces of hundreds of pN may appear
unreasonable given that typical critical forces and torques for disrupting
B-DNA structure in single--molecule experiments are only $f_{c}\simeq65\,\text{pN}$
and $t_{c}\simeq40\,\text{pN\, nm}$ \citep{sarkar01}, and that unzipping
occurs already at $\simeq15\,\text{pN}$. To see that there is in
fact no contradiction, note the qualitative difference between protein
forces acting locally on B-DNA, and externally applied micro-manipulation
forces. Consider again the stereotypical double-well free energy in
Fig.~\ref{fig:doublewell}, where the potential wells at $x_{1}$
and $x_{2}$ could for example correspond to B-DNA and to an overstretched
state (S-DNA), respectively. Without external force, the ground state
is $x_{1}$. A protein which binds DNA constrains the base-pair step
to a position $x_{d}$ close to $x_{1}$. The required mean force
is $f_{d}=A'(x_{d})$; it is entirely determined by the local potential
well. On the other hand, an external force $f$ pulling on the DNA
fragment acts by tilting the free energy landscape, such that eventually
$x_{2}$ becomes the ground state. The critical external force $f_{c}=\Delta A/\Delta x$
for the transition is determined by the free energy difference between
the potential wells and their separation, independent of local well
steepness. Thus for steep but nearly degenerate and well separated
potential minima, one gets $f_{d}>f_{c}$ for moderate displacements
$x_{d}-x_{1}$. Entropic effects have been neglected; they add minor
corrections.

In the example of the overstretching transition, the critical force
is $f_{c}\simeq65\,\text{pN}$ \citep{sarkar01}. The stiffness and
thermal standard deviation of base-pair elongation are $k_{1}\simeq10\kT/\text{\AA}^{2}$
and $0.33\,\text{\AA}$, respectively \citep{becker07}. Thus, stretching
the base-pair step by one standard deviation already requires twice
the critical force (and $4f_{c}$ if the next step is compressed by
the same amount).

\subsection*{Sharp elastic energy peaks result from balanced pairs of force and
torque}

The elastic energy gives a scalar measure of the overall deviation
of each base-pair step from its equilibrium conformation. Elastic
energy profiles can therefore quickly show the hot spots of local
deformation of complexed DNA. A recurring motif in the analyzed profiles
is an isolated high energy peak flanked by low energy steps. The energy
peak motif generally indicates a force and torque \emph{pair} deforming
the high-energy step, which approximately satisfies a local force
and torque balance. An example is provided by base-pair step 11-12
in the lac repressor (Fig.~\ref{fig:nmr}). The three-dimensional
force and torque vectors show that this step is kinked by opposing
pairs of force and torque with stretching, shearing and underwinding
components. Further examples with less complete balance is presented
by the stretching force pair at base-pairs 9 and 10 in TBP (Fig.~\ref{fig:tbp}),
and by the more weakly deformed, symmetry-related step 15-16.

\subsection*{Directions of deformation and force are non-trivially related}

Force and torque vectors often do not point into the directions one
would expect when picturing DNA as made up from some uniform isotropic
elastic material. This non-intuitive feature is visible in Fig.~\ref{fig:portraits}
but occurs also in force analyses of complex structures.

For example, regarding the forces needed to produce the $80^{\circ}$
turn in TBP (Fig.~\ref{fig:tbp}) one may have two non-exclusive
naive expectations: two point forces could push the ends of the curved
region at base-pairs 9 and 16 towards the center of the circle of
curvature, compensated by a force pulling the center of the curved
region away from this point; or distributed torques along the curved
region could try to bend DNA into the major groove, their torque vectors
pointing normal to the local plane of bending, i.e.~towards one of
the backbones. The computed forces and torques prove both of these
expectations wrong. Clearly all forces observed in the complex point
roughly along the local helical axis, not perpendicular to it; and
distributed torques do occur but point into the major groove, at right
angles to the expected direction. Thus, the coupled mechanical properties
of DNA produce the observed $80^{\circ}$ turn of the TATA box by
an array of torques as would result, from pulling both sugar-phosphate
backbones into the respective 3' direction. Interestingly, in an MD
simulation of a TATA sequence without protein \citep{flatters97},
this mode of external 3'-pulling was observed to produce a bent shape
that mimics the bound conformation of DNA in the complex.

Another example of this coupling is the $36^{\circ}$ Roll of the
base-pair step 9-10 of the lac repressor complex. As can be seen in
Fig.~\ref{fig:nmr}, it results not from a bending torque but mainly
from stretching and underwinding by the protein. In summary, the non-trivial
nano-mechanical properties of DNA make it impossible to tell by eye
what force and torque directions are required for a particular shape.

\subsection*{Forces require DNA-protein contacts, but contacts do not always transmit
force}

In the absence of long-range interactions, only base-pairs that are
contacted by protein should ever experience external forces. The converse
is not true: not every contact can be expected to actually transmit
force. These requirements  allow for a consistency check of a DNA
nano-mechanics analysis by comparing calculated external forces with
observed contact points. For example, in the lac repressor, base-pairs
2, 3, 21 and 22 are non-contacted, and indeed their forces and torques
are weak. The remaining magnitude gives an estimate of the error in
force determination of about 10\% of the peak force. This error estimate
agrees with the standard deviation of forces and torques across the
NMR ensemble, cf.~Fig.~\ref{fig:nmr}. We conclude that possible
systematic errors in force determination are smaller than the uncertainty
due to limited structural precision. In contrast, the calculated forces
and torques at the non-contacted end base-pairs 1 and 23 of the binding
site are about three times the error estimate. Their non-zero forces
point to a systematic error, possibly due to the dissimilar properties
of internal and end base-pairs, see Methods. Base-pair 9 in the same
complex is an example of a base-pair in close contact with protein
residues which experiences forces indistinguishable from 0, showing
that contacts do not imply local forcing.

A similar situation can be found in the TBP repressor complex \citep{bleichenbacher03},
Fig.~\ref{fig:tbp}. Here the boundary base-pairs 1 and 2 are contacted
by protein from a neighboring unit cell, not shown in Fig.~\ref{fig:tbp},
and are therefore not expected to be free. In contrast base-pairs
3 to 7 represent a stretch of suspended, non-contacted poly-G DNA
and are expected to be force-free. Indeed their residual force and
torque magnitudes are only about 20\% of the thermal force scale;
this margin can serve as an error estimate.

\subsection*{DNA is deformed by a combination of local forces and propagated tension}

 Apart from local forces, tension from flanking DNA can deform a
base-pair step, see Fig.~\ref{fig:portraits}. A well-known extreme
example of tension propagation is the lac operon, where a tight loop
of many base-pairs is held together by two copies of the lac repressor
dimer \citep{mossing86,kramer87,zhang06a,swigon06}. While forces
are exerted only at the ends, the propagated tension deforms DNA along
the loop. In protein-DNA complexes, the observed deformations of bound
DNA are generally due to the combination of local external forces
and torques and distant forces and torques, propagated as tension
along the chain. This non-local part is always present when forces
do not balance locally, but becomes most apparent when deformations
occur without local forces. Considering base-pairs 7-9 in the I\emph{ppo}-I
complex, Fig.~\ref{fig:ippo}, note that both steps 7-8 and 8-9 are
stressed as indicated by their high elastic energy. However base-pair
8 in the middle experiences only weak external force. This motif can
be interpreted mechanically as follows: The protein pulls base-pairs
7 and 9 apart by a nearly antiparallel force-pair leaving base-pair
8 suspended freely in the middle. Interestingly, the single strand
cuts performed by the functional form of the I\emph{ppo}-I complex
occur exactly at the pre-stretched base-pair step 8-9 and the symmetry-related
site 12-13.

\section*{Conclusions}

In this article we have presented an analysis of DNA nano-mechanics
within a given complex structure as a novel but natural way to think
about the interaction of DNA with proteins. The free energy function
of any coarse-grained model allows calculation of the mean forces
acting on the represented degrees of freedom. Our basic idea is to
infer these mean forces from structural data and use them to describe
the mechanical interaction of DNA with its environment. This gives
an intuitive way to interpret the mechanics of DNA-protein binding,
augmenting the interpretation of molecular conformation. 

We have implemented this idea for the rigid base-pair model of DNA,
using an efficient and compact matrix formalism to derive the forces
and torques acting on base-pairs. This level of coarse-graining offers
good compromise between resolution and reliability. In particular,
the available parameter sets summarize comprehensive structure database
analysis and large-scale MD simulation efforts. Our method puts this
large body of DNA elasticity to work in a computationally inexpensive
way. As we have demonstrated, the results are robust with respect
to differences in crystallization and among conformers in an NMR structure
ensemble, force profiles using different parameter sets converge after
pre-relaxation, and forces on non-contacted base-pairs vanish within
the estimated error bounds. New, nonlinear or poly-nucleotide potentials
will lead to improved accuracy and range of applicability of nano-mechanics
analysis.

From a physical chemistry point of view, base-pair forces and torques
are interesting in their own right since they describe the inter-molecular
force balance. They are easy to interpret since they are local quantities.
However, they are \emph{not} easy to deduce from a structure `by eye',
since they depend non-locally on DNA deformation, by propagation of
elastic tension. Thus while it only takes a crystal structure as input,
DNA nano-mechanics analysis can improve on pure conformation analysis
by integrating prior knowledge about DNA elasticity.

To predict indirect readout effects, i.e.~sequence specificity of
proteins mediated by DNA elasticity and structure, it is desirable
to calculate the elastic contributions to protein-DNA binding free
energies, see e.g.~\citep{becker06}. Nano-mechanics analysis is
not directly useful for this purpose, since energies can be calculated
directly from the deformations. However, good estimates for the total
elastic energy of a protein-DNA complex require some elastic model
of the protein. Comparison of predicted and structure-base base-pair
forces appears a good way to validate such coarse-grained protein
models.

A related application of our method is to establish a connection between
multi-scale bio\-mole\-cular simulation and experimental structural
data. Common simulation schemes connect different levels of coarse-graining
by force-matching \citep{izvekov05,ayton07}. Simulated and structure-based
base-pair forces can be matched with little extra effort, suggesting
a data-driven method for the rational design and validation of new
coarse-grained protein models. Our analysis also establishes a link
between structural studies and biophysical force measurements on short
DNA loops \citep{shroff08}. 

From a biochemistry point of view, interpretation of structures in
terms of interaction forces leads to hypotheses about their biological
functioning. For instance, DNA-modifying proteins such as nucleases
inflict strong DNA deformations; when trapped intermediate states
can be crystallized, their base-pair interaction forces shed light
on the reaction mechanism, see the \emph{Ippo-}I example in Results.
Furthermore, the strength of transmitted force constitutes a measure
to classify local sites of interaction, as applied in a related article
on nucleosome nano-mechanics \citep{nuc}. The strongest--force contacts
play the most important role in enforcing the structural constraints
implied by binding. Thus, mutation of a DNA base or a protein residue
affecting a high-force contact site is expected to result in strong
perturbation of the complex structure, while small-force contacts
should only weakly affect the global structure of the complex. So
whenever the global DNA conformation on a scale of several base-pairs
is relevant for biological function of a complex, high-force contact
sites emerge as natural targets for mutation assays. We refer the
reader to \citep{nuc} for a first observation of the effect of mutations
on the force patterns.

\section*{Acknowledgment}

This work was supported by the chair of excellence program of the
Agence Nationale de la Recherche (ANR). The authors wish to thank
R. Lavery for helpful discussion.

\clearpage{}

\bibliographystyle{unsrtnat}
\bibliography{long.1}
\onecolumn


\section*{Supplementary figures and text}

\floatname{sfigure}{Figure}
\newfloat{sfigure}{tbp}{sfigs} 
\renewcommand{\thesfigure}{\mbox{{\it{}supp-}\arabic{sfigure}}}

\renewcommand{\thepage}{}
\renewcommand{\theequation}{\mbox{{\it{}supp-}\arabic{equation}}}
\renewcommand{\thesubsubsection}{\mbox{Text~{\it{}supp-}\arabic{subsubsection}}}
\setcounter{equation}{0}
\setcounter{subsubsection}{0}

\newpage{}

\begin{sfigure}

\centering{\includegraphics[width=3.25in]{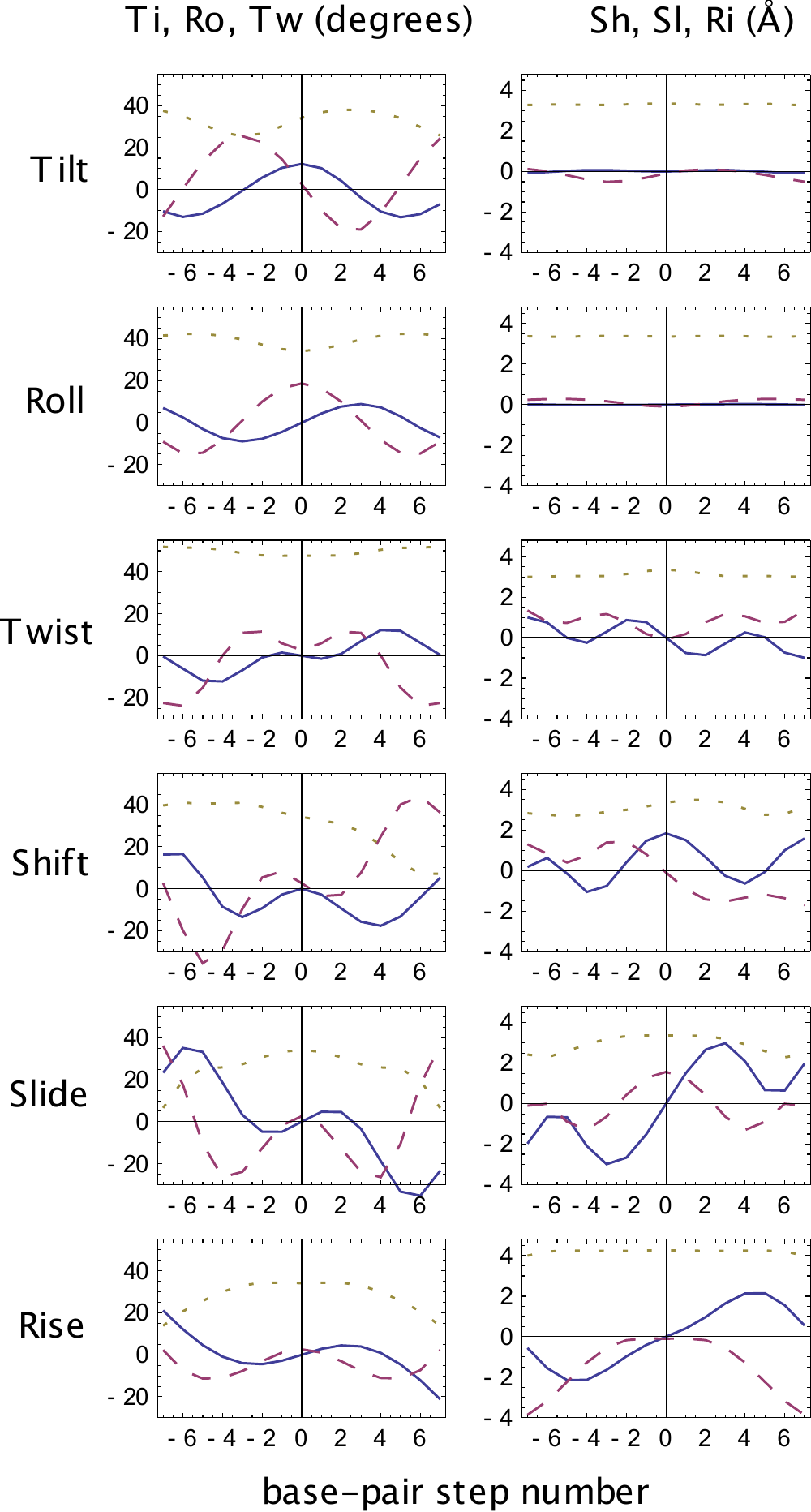}}

\caption{\label{fig:portplots} Tilt, Roll, Twist (left column) and
Shift, Slide, Rise (right column) base-pair parameters corresponding
to the equilibrium shapes of homogeneous DNA shown in the rightmost
column of Fig.~\ref{fig:portraits}. Tilt and Shift, solid line;
Roll and Slide, dashed; Twist and Rise, dotted. At the central base-pair
step 0, all parameters except for the perturbed one, are at their
equilibrium values in each row. One can see that unlike the case of
a shearable rod with uncoupled modes and isotropic bending \citep{shi96},
the excess twist is \emph{not} constant. As a consequence, external
forces acting in non-equilibrium shapes cannot be deduced directly
from local excess base-pair parameter values. Due to DNA symmetry
properties, the two halves of the chain with positive and negative
bp are identical only in the Roll, Twist, Slide and Rise panels. In
these panels, the excess base-pair step parameters exhibit (anti)-symmetric
profiles. The elastic energy (not shown) also varies along the chain.}\end{sfigure}

\clearpage{}

\subsubsection{Base-pair step potentials in exponential coordinates\label{sub:expcoords}}

We parametrize a base-pair step conformation $\mathbf{g}_{kk+1}$
by letting\begin{equation}
\mathbf{g}_{kk+1}=\mathbf{g}_{\mathrm{eq},b_{k}b_{k+1}}\exp[q_{kk+1}^{j}\mathbf{X}_{j}]\end{equation}
where $\exp$ is the matrix exponential and $\mathbf{g}_{\mathrm{eq},b_{k}b_{k+1}}$
is the equilibrium base-pair step conformation. That is, step conformations
are given in exponential coordinates on the rigid motion group, based
at the point $\mathbf{g}_{\mathrm{eq},b_{k}b_{k+1}}$ \citep{becker07}.
The harmonic step energy function can be written as\begin{equation}
\fe_{bb'}(\mathbf{g}_{kk+1})=\tfrac{1}{2}q_{kk+1}^{i}\mathbf{S}_{(b_{k}b_{k+1})ij}q_{kk+1}^{j},\label{eq:quaden}\end{equation}
where $\mathbf{S}$ can be obtained from stiffness matrices given
in base-pair step parameters by multiplying with appropriate Jacobian
matrices. In the coordinates introduced above, the external forces
have simple expressions for weakly deformed steps. One obtains to
first order,\begin{equation}
\mu_{(k)i}=\mathbf{S}_{(b_{k-1}b_{k})ij}q_{k-1k}^{j}-(\mathbf{A}_{kk+1}^{\mathsf{T}}\mathbf{S}_{(b_{k-1}b_{k})})_{ij}q_{kk+1}^{j}+O(q)^{2},\label{eq:muexpcoord}\end{equation}
where $\mathbf{A}_{kk+1}=\mathbf{Ad}(\mathbf{g}_{\text{eq},b_{k}b_{k+1}})$
and $\mathbf{Ad}$ denotes the adjoint representation of the group;
for details, see \citep{becker07}. For base-pair steps that are deformed
more strongly, it was necessary to consider the corrections to this
first order result. We therefore postulated the quadratic energy Eq.~\ref{eq:quaden}
to be valid for finite extensions, and recovered $\mu_{(k)i}$ by
using the Jacobian matrix $\mathbf{J}_{\exp}$ relating exponential
coordinates to the left invariant frame; one then gets\begin{equation}
\mu_{(k)i}=(\mathbf{J}_{\exp}^{\mathsf{T}}(q_{k-1k})\mathbf{S}_{(b_{k-1}b_{k})})_{ij}q_{k-1k}^{j}-(\mathbf{A}_{kk+1}^{\mathsf{T}}\mathbf{J}_{\exp}^{\mathsf{T}}(q_{kk+1})\mathbf{S}_{(b_{k}b_{k+1})})_{ij}q_{kk+1}^{j}.\label{eq:muliframe}\end{equation}
Here, the Jacobian is given by $\mathbf{J}_{\exp}^{-1}(q)=\int_{0}^{1}\mathbf{Ad}(\exp(-s\, q^{i}\mathbf{X}_{i}))\,\dd s$.\clearpage{}

\begin{sfigure}\centering{\includegraphics[width=3.25in]{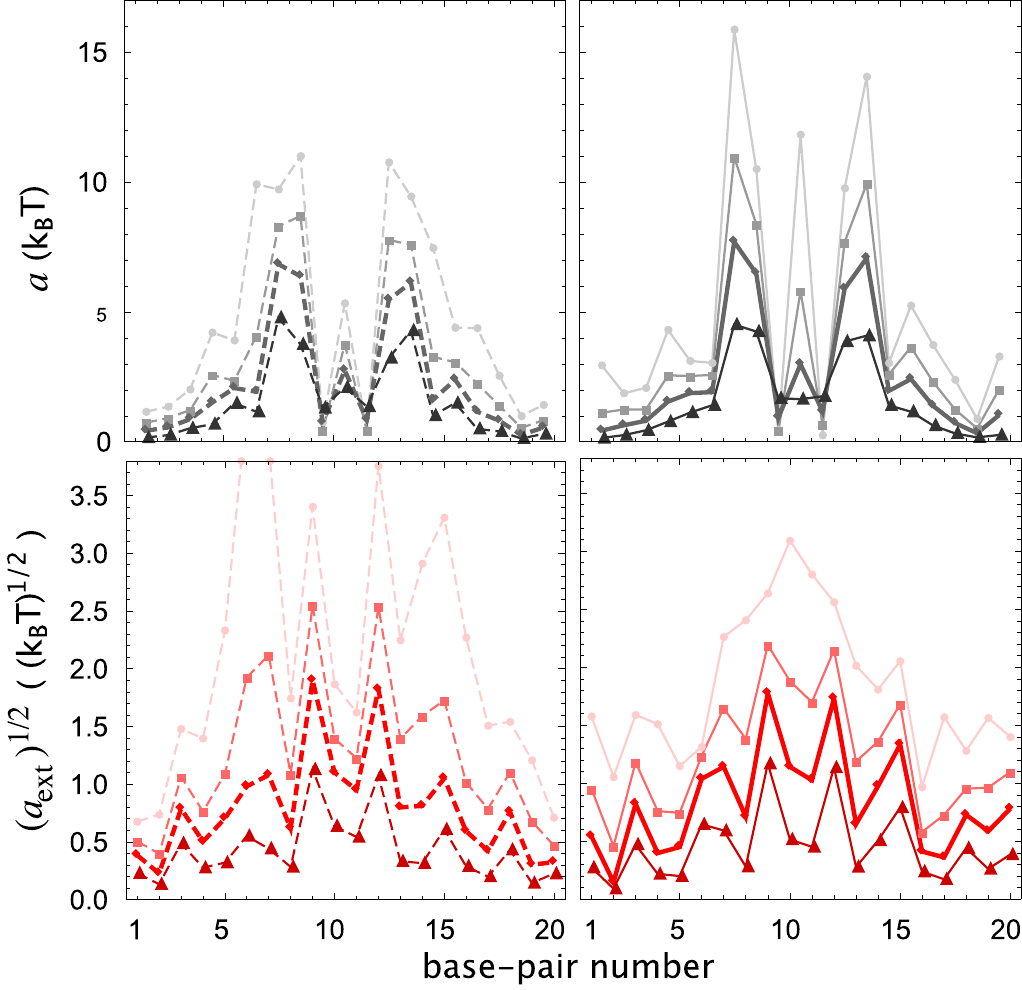}}

\caption{\label{fig:relax}The magnitude of base-pair forces and
torques resulting from nano-mechanics analysis depends on the range
of allowed pre-relaxation $r_{m}$. Generally, wider allowed relaxation
ranges result in a reduction of global force and energy scale, scaling
roughly as $r_{m}^{-1}$ . In addition, for the smaller values of
$r_{m}$ the profile shapes also change. The first row of the figure
shows energy profiles for I\emph{ppo}-I. The curves correspond to:
no relaxation and $r_{m}=0.15,0.3,0.6\mathring{\mathrm{A}}$, from
top to bottom in each panel. The value $r_{m}=0.3\mathring{\mathrm{A}}$
equals to the reported atomic position uncertainty, used in Fig.~\ref{fig:ippo}.
MP parameter set, left column; P parameter set, right column. The
second row shows the combined magnitude of external force and torque,
computed from the energy $\fe_{\text{ext}}(\mu_{(k)})$ associated
with a force-torque pair: $\fe_{\text{ext}}(\mu_{(k)})=\tfrac{1}{2}\mu_{(k)}^{\mathsf{T}}(\mathbf{S}_{(b_{k-1}b_{k})}+\mathbf{A}_{kk+1}^{\mathsf{T}}\mathbf{S}_{(b_{k}b_{k+1})}\mathbf{A}_{kk+1})^{-1}\mu_{(k)}.$
Values of $r_{m}$ and parameter sets as above.}\end{sfigure}
\end{document}